\documentclass[useAMS,usenatbib]{mn2e}
\usepackage{graphicx} 
\usepackage{amssymb}
\newcommand {\hi} {{\rm H}\,{\sc\rm I}}

\newcommand {\kms} {\,{\rm km\,s}^{-1}}

\newcommand {\erg} {\,{\rm erg}}
\newcommand {\pc} {\,{\rm pc}}
\newcommand {\cm} {\,{\rm cm}}
\newcommand {\kpc} {\,{\rm kpc}}

\newcommand {\msun}{\,{\rm M}_\odot}

\newcommand{\Myr}{\,{\rm Myr}}

\newcommand{\K}{\,{\rm K}}

\newcommand {\msunyr}{\,{{\rm M}_\odot\,\rm yr}^{-1}}

\title[The origin of the high-velocity cloud complex C]
{Galactic hail: The origin of the high-velocity cloud complex C}

\author[F. Fraternali, A. Marasco, L. Armillotta \& F. Marinacci]
{
	F. Fraternali$^{1, 2}$\thanks{E-mail: filippo.fraternali@unibo.it}
	A. Marasco$^{2}$
	L. Armillotta$^{1}$
	F. Marinacci$^{3, 4}$
	\\
	$^{1}$Department of Physics and Astronomy, University of Bologna, viale Berti Pichat 6/2, 40127, Bologna, Italy\\
	$^{2}$Kapteyn Astronomical Institute, Postbus 800, 9700 AV, Groningen, The Netherlands\\
	$^{3}$Heidelberger Institut f\"{u}r Theoretische Studien, Schloss-Wolfsbrunnenweg 35, 69118 Heidelberg, Germany\\
	$^{4}$Zentrum f\"{u}r Astronomie der Universit\"{a}t Heidelberg, Astronomisches Recheninstitut, M\"{o}nchhofstr. 12-14, 69120 Heidelberg, Germany\\
}
\begin{document}

\date{Accepted xxx Received xxx; in original form xxx}

\pagerange{\pageref{firstpage}--\pageref{lastpage}} \pubyear{2014}
\maketitle
\label{firstpage}

\begin{abstract}
High-velocity clouds consist of cold gas that appears to be raining down from the halo to the disc of the Milky Way. 
Over the past fifty years, two competing scenarios have attributed their origin either to gas accretion from outside the Galaxy or to circulation of gas from the Galactic disc powered by supernova feedback (galactic fountain). 
Here we show that both mechanisms are simultaneously at work.
We use a new galactic fountain model combined with high-resolution hydrodynamical simulations. 
We focus on the prototypical cloud complex C and show that it was produced by an explosion that occurred in the Cygnus-Outer spiral arm about 150 million years ago. 
The ejected material has triggered the condensation of a large portion of the circumgalactic medium and caused its subsequent accretion onto the disc.
This fountain-driven cooling of the lower Galactic corona provides the low-metallicity gas required by chemical evolution models of the Milky Way's disc.
\end{abstract}


\section{Introduction}

Disc galaxies like the Milky Way must continuously accrete low-metallicity gas from the intergalactic medium (IGM) in order to feed their star formation. 
Indirect evidence for gas accretion comes from estimates of the gas consumption times \citep{Bigiel+11}, star formation histories \citep{Fraternali&Tomassetti12} and chemical evolution models \citep{Schoenrich&Binney09}. 
However, gas accretion is very difficult to observe directly and the way it takes place remains unknown \citep{Sancisi+08} and alternative scenarios have been proposed \citep{Leitner&Kravtsov11}.
According to cosmology, most of the ordinary matter in the Universe resides in the IGM and it is largely unobserved. 
A fraction of these so-called ``missing baryons'' are expected to be in the form of a hot and diffuse gas that reaches its highest densities around galaxies and forms extended cosmological coronae \citep{Fukugita&Peebles06}. 
Recent observations indicate that the masses of these coronae are comparable to the total mass in galaxy discs \citep{Dai+12,Gatto+13}.
Thus, galaxies have potentially large reservoirs for gas accretion.

High Velocity Clouds (HVCs) are cold gas complexes having velocities incompatible with their being part of the Galactic disc.
Since their discovery, two competing scenarios have been proposed suggesting origins either external or internal to the Milky Way's disc \citep{Wakker&vanWoerden97}. 
The external origin was originally put forward by \citet{Oort70} who interpreted them as left-overs from the formation of our Galaxy. 
The internal hypothesis postulated that HVCs were clouds condensing in the Galactic halo from gas previously ejected from the disc by supernova explosions, through a galactic fountain \citep{Bregman80}.
The measurement of distances and metallicities of the HVCs became crucial to distinguish between the two possibilities.

\citet{Wakker+99} measured the metallicity of the most prominent and best known of all HVCs, complex C, to be significantly below the Solar value.
In subsequent years, metallicities have been measured for most complexes and they have often been found to be below the Solar value, typically between 0.1 and 0.5 Solar \citep{Wakker01}.
Moreover, the distances to the major complexes have all been found to be within about 10 kpc from the Galactic plane \citep{Wakker+08}.
Overall, the relatively low metallicity was considered evidence for an extragalactic origin although the actual source of the HVC material remained a mystery.
Various mechanisms have been proposed such as thermal instabilities in the Galactic corona \citep{Kaufmann+06}, gas stripping from satellites \citep{Olano08} and gas condensation in cosmological filaments \citep{Fernandez+12}.

In this Letter, we present a model that explains the formation of Complex C from a mixture of internal and external mechanisms, as the ejection of disc material from supernova feedback triggers the cooling of a large portion of the Galactic corona.

\section{Method}

\begin{figure}
\begin{center}
\includegraphics[width=0.4\textwidth]{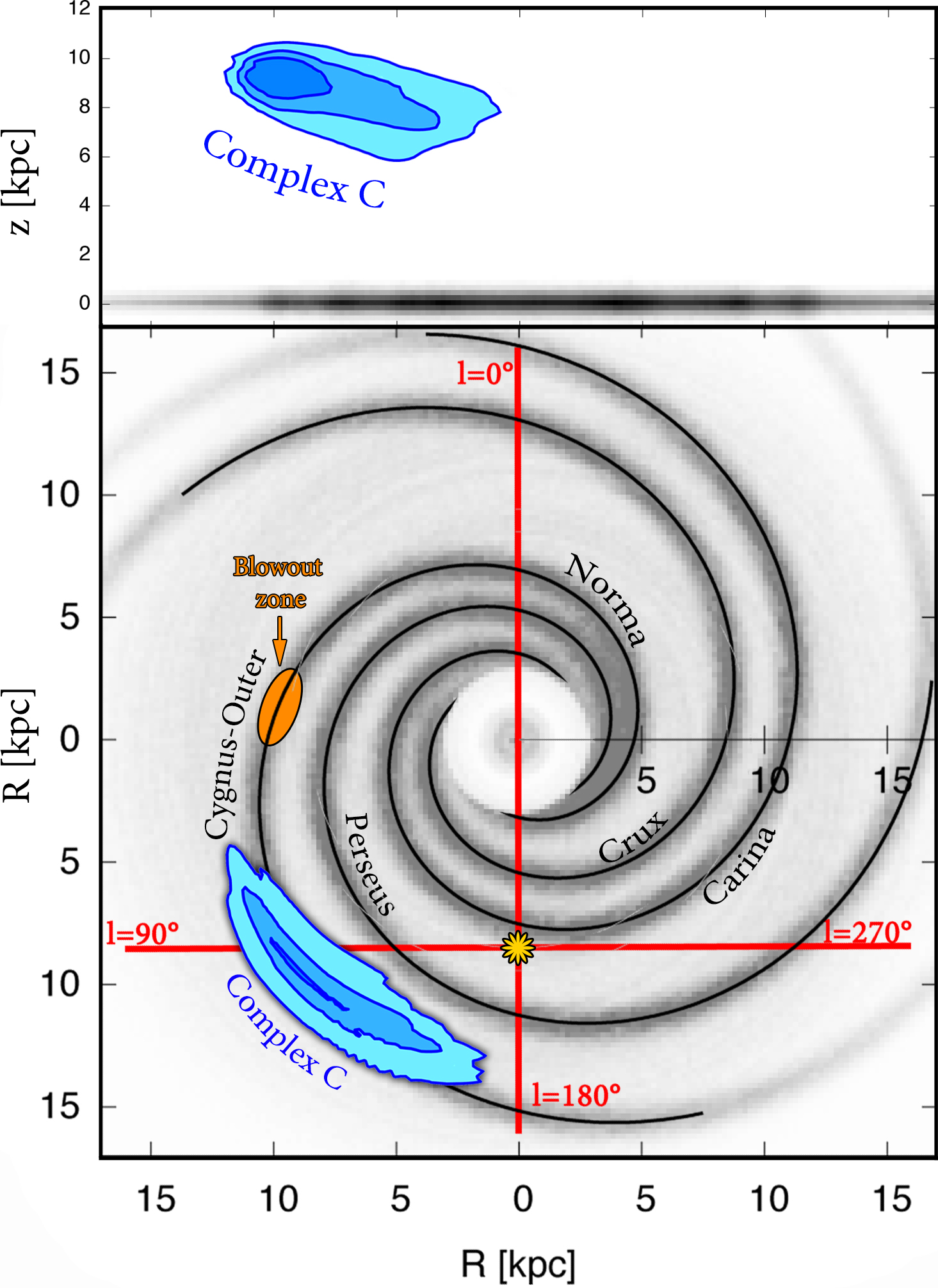}
\caption{Location of HVC complex C seen from two viewpoints. 
Top panel: edge-on projection seen from the anticentre. 
Bottom panel: face-on view; the two panels are on the same scale. Contours are obtained from our galactic fountain model and correspond to $\hi$ column densities of 1, 2 and $3\times 10^{19} \cm$. 
The spiral arms, the position of the Sun and the location of the ejection (orange ellipse) of the seed material are shown.
\label{fig:overview}}
\end{center}
\end{figure}

We use a combination of a galactic fountain dynamical model and high-resolution hydrodynamical simulations to model how disk gas, ejected outside the plane by stellar feedback, travels through the Galactic halo interacting and mixing with the circumgalactic medium (the Galactic corona).
This mixing dramatically reduces the cooling time of the corona causing its condensation into structures that are then pulled down to the disc by gravity \citep{Marinacci+10}, in analogy to rain and hail in the Earth's atmosphere. 
An axi-symmetric galactic fountain model of this kind produces an excellent fit to the all-sky extraplanar $\hi$ emission of the Milky Way \citep{Marasco+12}. 
A crucial modification that we made to this model is to assume that stellar feedback preferentially ejects material in the regions of the spiral arms \citep[see also][]{Struck&Smith09}.
We adopted a four-arm spiral pattern model \citep{Steiman-Cameron+10} and an arm width (FWHM) of 1 kpc. 
We assumed a pattern speed of $25 \kms$ \citep{Gerhard11}, which implies that the Sun is located roughly at the co-rotation radius, assuming $R=8.5 \kpc$ and $v_{\odot}=220 \kms$ (see Fig.\ \ref{fig:overview}).

The orbits of the fountain particles are integrated in a standard Galactic potential, Model I in \citet{Binney&Tremaine08}, starting from initial conditions that can be varied to fit the data. 
The trajectories calculated for gravity only are modified by two hydrodynamical effects: ram pressure (drag) and condensation of coronal gas \citep[see][for details]{Marasco+12}.
The drag time is a function of the mass and cross-section of the cloud \citep{Fraternali&Binney08}.
We assume that complex C is made up of tens of clouds with masses of $\sim 10^5 \msun$ that are shredded and disrupted by turbulence giving rise to the observed power-law mass function \citep{Hsu+11}. 
This assumption fixes the drag time, however experiments with different initial masses yield very similar final results.

To determine the condensation of the coronal gas we ran a series of 2D hydrodynamical simulations of fountain clouds traveling through the Galactic corona. 
The resolution required to accurately characterize the condensation process is at least $2\times2 \pc$ \citep{Marinacci+10}.

\begin{figure}
\begin{center}
\includegraphics[width=0.45\textwidth]{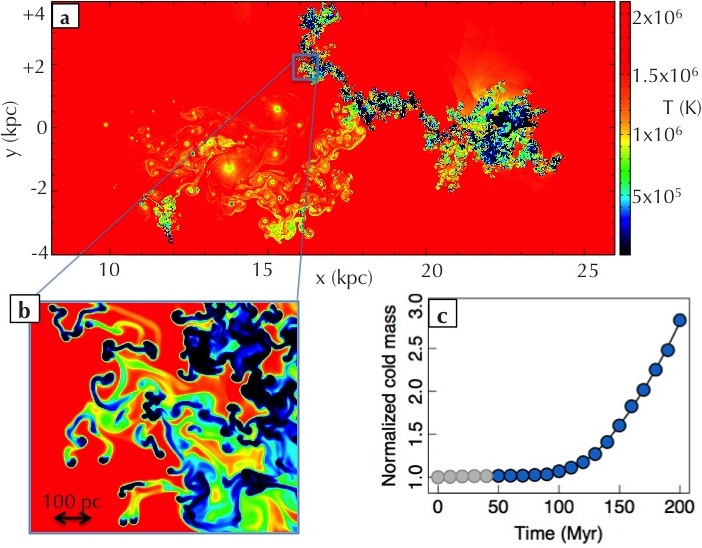}
\caption{
Hydrodynamical simulations of a galactic fountain cloud traveling through the Galactic corona.
The simulation starts (at $x=0$) as a round cloud at nearly Solar metallicity, moving from left to right with respect to the hot metal-poor gas of the Galactic corona. a) Temperature map after 175 Myr;
the cold (blue) gas in the upper snapshot is made up both by cloud gas and by gas condensing from the corona, see close-up (b).
c) Mass of cold (below $10^{4.2} \K$) gas as a function of time normalized to the initial mass of the cloud; the evolution before $t_{\rm delay}=46 \Myr$ is determined by the unrealistic initial configuration of the cloud and has been excluded.
}
\label{fig:sims}
\end{center}
\end{figure}

We consider a cold ($10^4 \K$) cloud at nearly Solar-metallicity travelling through a hot medium with an initial velocity of $180 \kms$, which produces relative velocities with respect to the corona compatible with those determined by our dynamical model. 
The corona is at $T=2\times10^6 \K$, roughly the virial temperature of the Milky Way \citep{Fukugita&Peebles06} and at a metallicity $Z=0.1$ Solar \citep{Hodges-Kluck&Bregman13}. 
The simulation runs for 200 Myr during which the cloud is torn apart by hydrodynamical instabilities and produces a long turbulent wake where disc and coronal material mix efficiently. 
In Fig.\ \ref{fig:sims}-a, we show a temperature snapshot after 175 Myr. 
The hot and rarefied atmosphere of the Milky Way's halo would normally have a cooling time exceeding 2 Gyr. 
However, the mixing with cold and metal-rich material reduces this time by orders of magnitude and condensation knots form in the cloud's wake (see close-up in Fig. 2-b). 
Fig.\ \ref{fig:sims}-c shows the evolution of the cold gas ($T< 10^{4.2} \K$) in our simulation: the cold gas mass increases to more than twice the initial mass of the cloud. 

We used the curve in Fig.\ \ref{fig:sims}-c to describe the condensation from the corona in the particles of our galactic fountain model.
This condensation, together with the effect of ram pressure modifies the trajectories of the particles, a modification that is directly observed in the $\hi$ data \citep{Fraternali&Binney08}. 
Given that our simulation starts from the unrealistic configuration of a round cloud we have to introduce a delay time ($t_{\rm delay}$) that allows us to exclude the initial few tens of Myr of the simulation \citep[see also][]{Marasco+12}.

The main observational constraint of our model is the LAB 21-cm all-sky survey of the Milky Way \citep{Kalberla+05}, which provides detailed information about $\hi$ column densities as a function of position in the sky and velocity along the line of sight. 
We created a mask of complex C by isolating its $\hi$ emission, channel by channel in the velocity range $-200< v_{\rm LSR}< -75 \kms$ (Fig. 3, blue shading). 
Our galactic fountain model predicts the $\hi$ emission from fountain clouds for different choices of our six free parameters: the initial velocity, the location along the spiral arm, the size of the ejection zone, the look-back time and the duration of the ejection and the delay time in the simulation.
The fit to the observations (minimization of residuals $|{\rm model}-{\rm data}|$) was performed in two steps: at first, we used a downhill simplex method to find the best-fit values; then, we ran a Monte Carlo Markov Chain (MCMC) to estimate the error bars on our parameters.
For details see the online material.

\section{Results}

\begin{figure*}
\begin{center}
\includegraphics[width=0.7\textwidth]{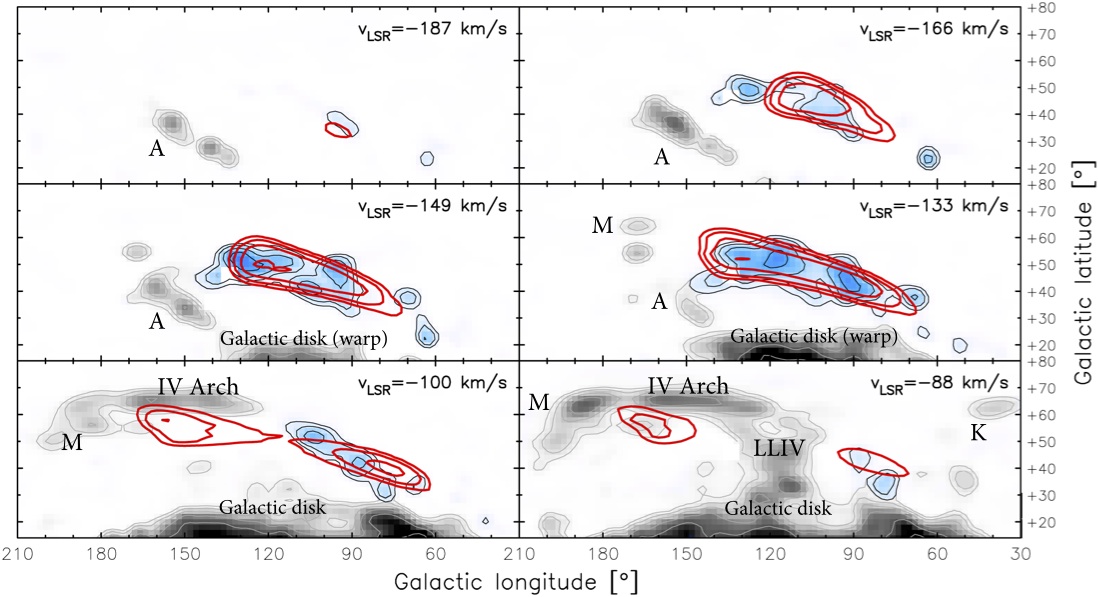}
\caption{
Six representative channel maps from the $\hi$ LAB survey (grey shade and contours) in the region of complex C (blue shade, black contours). The red contours show the $\hi$ emission predicted by our model. Both model and data are smoothed to $4^{\circ}$ of angular resolution. Note that our model interestingly predicts some emission at low velocities (around $90 \kms$), where there is $\hi$ detected although it is not usually associated to complex C. Contour levels are at 0.04, 0.08, 0.16, 0.32, 0.64 and $1.28 \K$. In this projection the ejection of seed material took place at about $l=45^{\circ}$ and $b=0^{\circ}$.}
\label{fig:channels}
\end{center}
\end{figure*}

We found that the only spiral arm that can be responsible for the origin of complex C is the (Norma)-Cygnus-Outer arm (Fig.\ \ref{fig:overview}). 
The best fit to the observations is obtained by ejecting material from this arm at $R=9.5 \kpc$ from the Galactic centre ($13.3 \kpc$ from the Sun) around longitude $l=45^{\circ}$. 
Fig.\ \ref{fig:overview} sketches the ejection area and the current location of complex C.
The quality of the fit to the observations can be appreciated in Fig.\ \ref{fig:channels} where we show six representative channel maps from the LAB survey in the region of complex C overlaid with the emission predicted by our galactic fountain model. 
Clearly, there is a remarkable agreement in location, velocity and shape between model and data. We have marked the emission from the disc of the Milky Way, the Galactic warp, HVC complexes A, M and K, and the IV arch, a prominent intermediate-velocity cloud located at $d< 3.5 \kpc$ from us with Solar metallicity \citep{Richter+01}, and completely unrelated to complex C.

Table 1 summarizes the physical properties of complex C as derived from our dynamical/hydrodynamical model.
The material has been ejected vertically at $v_{\rm 0}\sim 210 \kms$ as the result of the blowout of one or more superbubbles. 
The ejection took place 150 Myr ago, lasted $\sim50$ Myr and ejected $3.4 \times 10^6 \msun$ of hydrogen from the disc plane. 
The ejected gas acted as a seed for the condensation of the galactic corona bringing the total $\hi$ mass budget of the system to $\sim 7 \times 10^6 \msun$, in agreement with the estimated mass of complex C \citep{Wakker+07}. 
Presently, complex C is a structure that is 15 kpc in length, and stretches from close to the ejection zone up to a maximum distance from the plane of $\sim10 \kpc$ (Fig.\ \ref{fig:overview}). 
The average positions are $R=12.9 \kpc$ and $z=8.4 \kpc$. 
The distance from the ejection location is due to a combination of the typical trajectory of fountain clouds, which tend to move the gas outwards, the differential rotation of the disc and the pattern speed of the arms. 
The location and the elongation are remarkably similar to the outcome of hydrodynamical simulations of superbubble blowout in our Galaxy \citep{Melioli+09}. 

\begin{table*}
\caption{Physical parameters of the superbubble blowout that generated complex C} 
  \begin{tabular}{lcccccccc}
  \hline
& $v_{\rm 0}$& $R_{\rm 0}$ &	$\delta_{\rm arm}$ & $t_{\rm 0}$ & $\Delta_{\rm t}$ & $t_{\rm delay}$ & $M_{\rm HI}$ (seed) & $M_{\rm HI}$ (final)\\
& ($\kms$) &	(kpc)	& (kpc)	& (Myr)	&(Myr)	&(Myr)	&($10^6\msun$)	&($10^6\msun$)	\\
  \hline
Best-fit &	211	&9.5	&2.9	&150	&53	&46	&3.4	&6.8\\
MCMC $2-\sigma$ range &$203-216$	&$9.4-9.6$ &$2.6-3.7$ &$145-152$ &$49-56$ &$43-52$ &$3.0-3.8$ &$6.2-7.5$\\
  \hline
\end{tabular}

\smallskip
$v_{\rm 0}$: initial ejection velocity of the disc material; $R_{\rm 0}$ and $\delta_{\rm arm}$: Galactic radius and size of the ejection region (blowout) along the arm; $t_{\rm 0}$ and $\Delta_{\rm t}$: look-back time and duration of the ejection; $t_{\rm delay}$: delay time in the hydrodynamical simulation (see text and Fig. 2-c). $M_{\rm HI}$: $\hi$ mass of the seed (disc) material and total current mass of complex C after the condensation of coronal gas.\\
\end{table*}

\begin{figure}
\begin{center}
\includegraphics[width=0.23\textwidth]{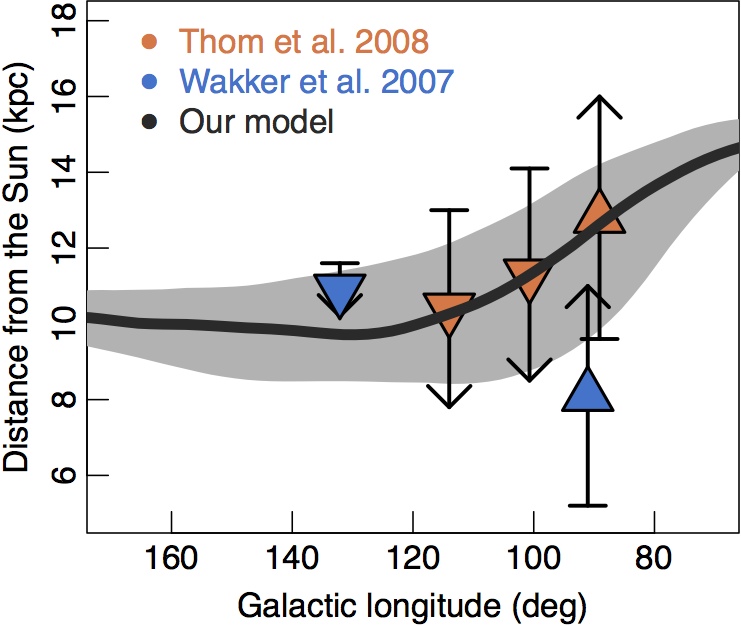}
\includegraphics[width=0.23\textwidth]{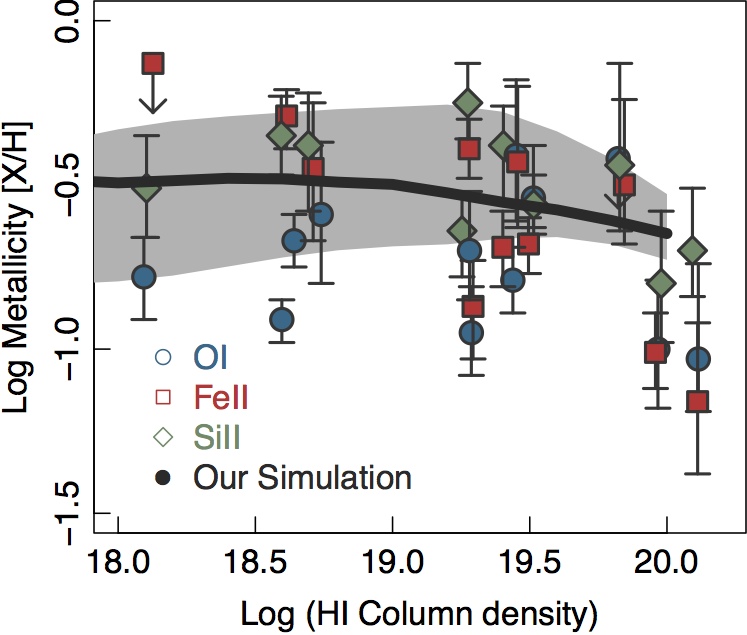}
\caption{
Left: average distance from the Sun of Complex C predicted by our dynamical model as a function of Galactic longitude (black curve), compared with distance determinations from the literature. Right: abundances of oxygen, iron and silicon as a function of $\hi$ column density compared to the prediction of our hydrodynamical simulation (black curve). The gray bands in both plots show the 16 and 84 percentiles.}
\label{fig:dist&Z}
\end{center}
\end{figure}

Having a model that reproduces the $\hi$ observations of complex C, we can compare its predictions to the other two available independent constraints, namely distances and metallicities. 
In the literature, there are several limits to the distance of complex C obtained from detections and non-detections of metal absorption lines against halo stars. 
The average distance is $10\pm2.5 \kpc$ \citep{Thom+08} with a mild gradient along the cloud. 
In Fig.\ \ref{fig:dist&Z} (left) we show the distance limits as a function of Galactic longitude compared to the average location of the gas in our dynamical model. 
In our model, the section of complex C closer to us is that at higher $l$ and $b$ (further away from the disc) and this is consistent with all the distance determinations. 
This agreement is highly significant given that the distance information has not been used in the fit to the $\hi$ data.

Our hydrodynamical simulation starts with $Z=0.1$ Solar for the coronal gas and $Z=0.8$ Solar for the Galactic ISM at $R\sim 9.5 \kpc$ \citep{Esteban+13}. 
During the evolution, a mass of gas larger than the initial mass of the cloud condenses from the corona bringing the average metallicity to $Z\sim0.27$ Solar at 200 Myr. 
Fig.\ \ref{fig:dist&Z} (right) shows the comparison between the predictions of our simulation and the observed abundances of complex C as seen in three (most reliable) elements as a function of the $\hi$ column density \citep{Collins+07}. 
We considered the gas at $T<10^{4.3} \K$ (typical temperatures of these absorption features) and analyzed the simulation snapshots between 150 and 200 Myr. 
We estimated the column densities and abundances along random vertical lines across the simulation boxes. 
The predicted metallicities encompass most of the observed data points. 
As for the distance, the metallicity has not been used in the fit to the $\hi$ data, it is in fact determined by the amount of condensation required to reproduce the $\hi$ kinematics.

\section{Discussion and conclusions}

From our dynamical model and the hydrodynamical simulations, we conclude that the most massive Galactic HVC (complex C) originated from an ejection of material from the Milky Way's disc. 
This material has traveled through the halo region, mixing with the hot gas from the Galactic corona and triggering its condensation. 
The source of the gas ejection must have been a prominent star forming region in the Cygnus-Outer arm at $l=45^{\circ}$, $b\sim 0^{\circ}$ and at $d\sim13.3 \kpc$ from us. 
The exact region cannot be easily recognized today as 
dust extinction along the line of sight prevents us seeing young open clusters in that area.

The development and blowout of a superbubble around a stellar association is the likely cause of the ejection \citep{MacLow+89}. 
We estimated an ejected hydrogen mass of $3.4 \times 10^6 \msun$ at a velocity of $\sim 210 \kms$. 
When correcting for He content, this corresponds to a total kinetic energy of $2\times 10^{54} \erg$, which can be achieved by $\sim 1 \times 10^4$ supernovae assuming an efficiency of 20\%. 
This would require the formation of $1-2\times10^6 \msun$ (depending on the IMF) of stars in 50 Myr in a region of a few $\kpc^{2}$. The inferred star-formation rate density is $\Sigma_{\rm SFR}\sim0.01 \msunyr \kpc^{-2}$, quite typical for a star-forming region in the Galactic disc.

A well known recent (30 Myr old) superbubble in the Milky Way is that in Ophiucus, located at $R\sim4 \kpc$ from the Galactic centre and currently ejecting material up to 3.5 kpc from the plane \citep{Pidopryhora+07}. 
We have used our galactic fountain model to estimate the ejection velocity of this material and found a value of $\sim 200 \kms$, as required for complex C. 
$\hi$ holes observed in the ISM of external galaxies are thought to be the remnants of superbubble blowouts. 
The kinetic energies associated with these holes are in the range $1\times10^{53}-1\times10^{55} \erg$ \citep[e.g.][]{Boomsma+08}. 
Thus, the ejection of the seed gas that caused the condensation of complex C is energetically plausible.

The cooling of the circumgalactic corona is a viable source for gas accretion in the Galaxy. 
However, the spontaneous development of thermal instabilities in the corona has been discussed and discarded in linear perturbation analyses \citep{Binney+09} and only non-linear perturbations are possible to explain HVC formation \citep{Joung+12}. 
The present model offers an elegant solution to this problem as the seeds for the instabilities are provided by the Galactic disc itself and the non-linearity of the flow is achieved by the development of turbulence. 
In this respect, HVCs should be regarded as the extreme manifestation of the galactic fountain phenomenon that is known to explain the origin of the intermediate-velocity clouds and the extraplanar gas in the Milky Way \citep{Marasco+12}. 
The intermediate temperature material predicted by the interaction with the corona (Fig.\ \ref{fig:sims}-a) is also consistent with the detection of ionized gas in the halo of our Galaxy \citep{Sembach+03,Lehner&Howk11,Fraternali+13}.

Predicting the amount of cold gas accretion contributed by complex C is not straightforward. 
Considering the most efficient scenario, in which the mass of condensed gas steadily increases as a power law up to $\sim350 \Myr$ (Fig. \ref{fig:sims}-c), we infer an upper limit of $M_{\rm accr}=0.25 \msunyr$ after correcting for the He content. 
This value is still well below what is needed to replace the gas consumed by star formation \citep{Chomiuk&Povich11}, thus confirming a scenario where most of coronal gas cooling is triggered by galactic fountains at the intermediate-velocity regime \citep{Marasco+12}. 
During the rare occasions when the ejection of material is extremely powerful (ejection velocities larger than $\sim150 \kms$) the gas will most likely appear in the high-velocity range. 
A long travel time through the halo will cause the condensation of a larger fraction of the corona and the metallicity will reach values between those of the disc and the corona.
The application of this scheme to the formation of all HVCs is beyond the purpose of this Letter, however our preliminary investigation shows that other complexes (e.g., complex K and L) can be explained in this way.
We conclude that supernova-driven cooling of the lower corona is a general mechanism for accretion of low-metallicity gas in disc galaxies as required by chemical evolution models.
It also explains why, at the present time, only star-forming galaxies acquire gas from their environment, while early-type galaxies have lost their ability to do so despite being surrounded by massive gas reservoirs \citep{Thom+12}.

\section*{Acknowledgements}
FF \& LA acknowledge support from PRIN-MIUR, prot. 2011SP-TACC. FM is supported by the DFG Research Centre SFB-881 ``The Milky Way System'' through project A1. AM acknowledges support from the European Research Council, 
Grant Agreement nr. 291531. The hydrodynamical simulations have been carried out at the CINECA FERMI IBM-Blue Gene/Q supercomputer (HP10C6D3E3) and at the Kapteyn Astronomical Institute (Gemini clusters).

\bibliographystyle{mn2e}
\bibliography{myBib}{}
\bsp

\label{lastpage}
\end{document}